\documentclass[a4paper,notitlepage]{article}
\usepackage{psfig, amsmath, amsfonts, amssymb}

\newcommand\mks{$\rm\mu s$}%
\newcommand\mkm{$\rm\mu m$}%

\begin{document}

\title{Daemon detection experiment}

\author{E. M. Drobyshevski\\{\it Ioffe Physical-Technical
Institute, Russian Academy of Sciences,}\\{\it 194021
St.Petersburg, Russia. E-mail: emdrob@pop.ioffe.rssi.ru} }

\date{}

\maketitle

\begin{abstract}
\noindent A month-long observation of two horizontal ZnS(Ag)
scintillating screens, 1 ${\rm m^2}$ in area and located one
above the other a certain distance apart, revealed about 10
correlated signals, whose time shift corresponds to an average
velocity of only $\sim$10--15 ${\rm km\,s^{-1}}$. We assign the
origin of these signals to the negative daemons, i.e.
electrically charged Planckian particles, which supposedly form a
part of the DM in the Galactic disk and were captured into the
near-Earth orbits. The estimated flux of daemons is
$\ge${}$10^{-4}\ {\rm m^{-2}s^{-1}}$. The key part in the
detection of daemons is played apparently by two processes: (i)
the daemon shedding the captured heavy nucleus in a few tens of
\mks\ as a result of a relatively rapid decay of the
daemon-containing nucleons, and (ii) emission of numerous Auger
electrons and nuclear particles occurring in the next capture or
recapture of a (heavier) nucleus by the daemon.
\end{abstract}

\section{Introduction. The daemon hypothesis}

Our Universe started from Planckian scales, and it appears only
natural to suggest that the major part of its mass, i.e., the DM,
is contained in primordial Planckian particles with 
$M = ({\rm\pi\hbar c/4G})^{1/2} \approx 2\times 10^{-5}$ g and
$r_{\rm g} = 2{\rm G}M/{\rm c^2} \approx 3\times 10^{-33}$ cm
\cite{mar65}. Such elementary black holes can be stable and
eternal \cite{bek94}. Multidimensional ($>$4) theories (e.g.
\cite{chan95}) allow the existence on them of stable electric
charge of up to $Z{\rm e} = {\rm G^{1/2}}M \approx 10{\rm e}$. We
assume that such DArk Electric Matter Objects, i.e. daemons,
carrying a charge of any sign (including possibly the zeroth one)
constitute a hierarchy of populations, viz. intergalactic,
Galactic halo (or crown), Galactic disk etc, with continuously
increasing concentration and decreasing mean random velocity. In
the disk, the DM makes up about 1/2 its total mass \cite{bah84}.
If it consists of daemons, their density here is 
$\sim\!10^{-12}\ {\rm m^{-3}}$. Due to their extremely low
concentration, negligibly small dimensions and a giant inertia,
the daemons, by themselves, constitute a non-collisional plasma
and, thus, do not capture and coalesce with each other. The
negative daemons' buildup inside the Sun is capable of accounting
for its energetics through catalysis of the proton fusion
reactions, and for the deficiency of the electron-capture
neutrinos \cite{dr96}.

Because of their large mass, daemons have a giant penetrating
power. That is why Markov \cite{mar65} was skeptical concerning a
possibility of detecting such particles. A charged daemon falling
from infinity onto the Sun acquires $\sim\!10^{23}$ eV, while
losing $10^{19}$--$10^{22}$ eV in passing along the Solar diameter
\cite{dr96}. Prior to becoming trapped, a daemon traverses many
times the Sun along gradually contracting, strongly elongated
orbits, whose perihelia lie within the Sun. If, in the course of
this process, the daemon enters the Earth's sphere of influence,
its path will change slightly, and the perihelion of the orbit
will leave the Sun with a high probability. This is how the
daemons can populate strongly elongated Earth-crossing
heliocentric orbits (SEECHOs). The flux of this population at the
Earth is estimated to be
$f_\oplus\sim 10^{-3}$--$10^{-6}\ {\rm m^{-2}s^{-1}}$ at 
$V \approx 35$--50 ${\rm km\,s^{-1}}$ \cite{dr97,dr00a}. Part of
this population will be gradually transferred into the near-Earth
almost-circular heliocentric orbits (NEACHOs), including those
crossing the Earth's surface. If the Sun moves relative to the
daemon population of the Galactic disk, these fluxes may undergo
seasonal variations at the Earth. The original goal of this work
was to detect the most dense (as we believed) SEECHO population.
In our opinion, such an approach should be more productive than
the standard searches for the strongly rarefied, Galactic halo DM
population (e.g. \cite{bra98,spo98}).

One may conceive the following consequences of the negative
daemon interaction with matter on the atomic and subnuclear
levels:

\noindent (1) When capturing (or recapturing) a nucleus, the
daemon first captures the ion of the atom, so that as it is
dropping to ever lower-lying levels, all the electrons of the
captured ion are emitted in the Auger process. Because the
binding energy of the daemon to the nucleus is measured in tens
of MeV, the energy of such Auger electrons may be as high as
$\sim$0.1--1 MeV;

\noindent (2) Catalysis of the fusion of light nuclei  
($Z_{\rm n}<Z$), including protons \cite{dr96,dr97}. As a result
of internal conversion, the energy of fusion in the vicinity of a
heavy charged particle is converted, as a rule, to the kinetic
energy of the resultant nucleus;

\noindent (3) Capture of heavy nuclei ($Z_{\rm n}\ge Z$). At 
$Z = 10$, it occurs in solid Fe, Zn, Sn in $\sim$2 \mkm\ at
$\sim$50 ${\rm km\,s^{-1}}$, and in $\sim$0.1 \mkm\ at $\sim$10
${\rm km\,s^{-1}}$ \cite{dr00a}. This `poisons' the catalytic
properties of the daemon. However, straightforward estimates
based on the solar energetics, if it is provided by
daemon-assisted catalysis of proton fusion, suggested that the
daemon should free itself of a captured heavy nucleus in
$\tau_{\rm ex} \approx 0.1$--1 \mks, apparently as a result of
the decay of the daemon-containing proton \cite{dr00a,dr00b} (for
$Z_{\rm n} > 24/Z$ the ground level of the daemon in a nucleus
lies inside one of its constituent nucleons). Note that the
uncertainty in the values of the parameters used in these
estimates permits one to shift $\tau_{\rm ex}$ in either
direction at least by an order of magnitude.

\noindent (4) If the `poisoned' daemon with $Z-Z_{\rm n}\ge 0$
encounters a heavier nucleus, due to great difference in binding
energies, it captures the latter while losing the previous one if
at the moment it is (or becomes) noticeably lighter than the anew
met nucleus (at $Z-Z_{\rm n}=0$ this `recapture' is an analog of
the charge exchange of ions moving through a neutral gas);

\noindent (5) Finally, the capture of a nucleus strongly excites
its internal degrees of freedom, which should stimulate the
emission of nuclear radiations (numerous nucleons and their
clusters, $\gamma$-quanta). Nuclei with a high $Z_{\rm n}$ have a
higher probability for the initiation of these processes.

Thus, all the above processes stimulate the emission of
radiations capable of producing scintillations, and this may be
used to detect the slowly moving daemons, whose direct impact
cannot generate a scintillation.

Earlier, we attempted to use the catalysis of light nucleus
fusion to detect the daemons. We employed an acoustic method with
two Li plates \cite{dr99}. The basic idea here was that the
material surrounding the trajectory should be rapidly heated by
the products of the ${\rm 2\,{^{7}Li} \rightarrow {^{14}C}}$
reaction. The thermal expansion of the material should generate a
sound wave, which then would be detected by piezoelectric
sensors. Their output signals would trace the daemon trajectory.
Our experiments revealed, however, an unexpectedly strong damping
of ultrasound in Li, and the experiments were stopped.

The measurements made with thick Be plates (4.5 cm), 0.12 
${\rm m^2}$ in area, coated on both sides with ZnS(Ag)
scintillator, were actually a continuation of the above
experiments along the same lines. It was assumed that the
products of the ${\rm 2\,{^{9}Be} \rightarrow {^{18}O}}$ reaction
escaping from thin near-surface regions, i.e. the points of
entrance into and egress from the Be plate of a daemon moving
along a SEECHO, would produce scintillations $\sim$1 \mks\ apart.
500 h of exposure did not yield any result \cite{dr00a}. The
latter experiment, however, prompted us to consider other
possible modes of daemon interaction with matter, particularly
the possible role of their `poisoning' by heavy nuclei [Fe, Si
impurities in Be, and the ZnS(Ag) nuclei].

Shedding the nucleus poisoning the daemon, combined with the
above-mentioned processes (1)--(5), suggest a variety of methods
for detection of slow daemons. Because of their large number,
both the Auger electrons and the radiations from heavy nuclei
excited in the capture are the most efficient in this respect.

\section{Description of the setup and assumed sequence of events
triggered by a daemon}

These considerations served as a basis when developing an
ideologically new and very simple setup of four modules.

Each module contains two parallel transparent polystyrene plates
4 mm thick and $50\times 50$ cm in size. The distance between the
plates was 7 cm. One of the polystyrene plate surface was coated
by type B3-s ZnS(Ag) powder $\approx 3.5\ {\rm mg\,cm^{-2}}$
thick. Its average grain size is 12 \mkm. The choice of this
classical phosphor, besides its availability, simplicity in use,
and a high light output, was motivated also by the fact that it
consists of medium-$Z_{\rm n}$ elements. And conversely, the
choice of polystyrene as a plate material was stimulated by its
low $Z_{\rm n}$, in order to reduce to a minimum the possibility
of heavy-nucleus poisoning of the daemon before its traversal of
the ZnS(Ag) layer. Each plate was viewed from a distance of 22 cm
by one FEU-167 PM tube with a dia. 100-mm photocathode. The
plates were separated by a sheet of black paper. To be able to
judge the essential features of the possible incoming daemon
flux, the ZnS(Ag)-coated surfaces of the polystyrene plates were
set facing the same side (down). As a result, the light entering
the top PM tube passed also a 4-mm thick transparent polystyrene
plate on the way. The polystyrene plates and the PM tubes viewing
them were placed in a cubic case 51 cm on the edge made of 0.3-mm
thick iron sheet with double-sided thin ($\approx$2 \mkm) facing
of tin. The upper horizontal case face was made of two sheets of
black paper. The PM photocathodes were arranged flush with the
horizontal case faces. All the four modules were placed side by
side in one horizontal plane. The total area of the four-module
detector was 1 ${\rm m^2}$.

The PM tubes were powered by a voltage corresponding to their
sensitivity of 10 ${\rm A\,lm^{-1}}$. A 4.5-k$\Omega$ resistance
served as a load. The signal from the load was supplied through a
1 mH inductance and a cable of total capacity 550 pF to an
oscilloscope. The purpose of this $L-C$ circuit was to stretch
the leading edge of the pulse so as to facilitate discrimination
of the long scintillations produced by heavy nonrelativistic
nuclei like $\alpha$-particles (Heavy-Particle Scintillations 
--- HPS; for their characterization the ${\rm ^{238}Pu}$
$\alpha$-source was used) against PM tube noise and short
scintillations caused by low-mass and relativistic particles
(cosmic rays etc) (the Noise-Like Scintillations --- NLS). Signals
from the two PM tubes of the same module were fed for comparison
into two inputs of one S9-8 dual-trace digital oscilloscope. The
latter was triggered by the output signal of the upper PM tube if
it reached a level $U_1\approx 2.5$ mV. The signal from the
second PM tube was considered significant if it increased in
0.4--1.5 \mks\ and its amplitude was $U_2 \ge 0.6$ mV. The
signals from the oscilloscopes recorded during $-$100 \mks\ before
the trigger (i.e. with a lead) and during $+$100 \mks\ after the
trigger (i.e. with a delay) were entered into computer memory if
they were seen in both traces.

The system was from the outset designed to detect daemons
impinging on it primarily from above. Indeed, it was assumed
originally that a daemon propagating through the polystyrene
plate from above would capture a heavy nucleus from the ZnS(Ag)
layer coating it on the back side, and emit in the process during
a certain time numerous Auger electrons and the excited nucleus
radiations which produce a protracted scintillation of HPS type.
If this was a daemon from the SEECHO population, it would, in
passing at a velocity $\approx\!35$--$50\ {\rm km\,s^{-1}}$ in
$\tau_{\rm ex}\sim 0.1$--1 \mks\ a path $\sim$0.3--5 cm, cause
decay of protons in the captured nucleus. Then the nucleus emits
products of the proton decay (pions etc.) and, possibly,
fragments of the nucleus itself. All of them also can give rise
to scintillation events, which also will be detected by the top
PM tube.

One can in principle conceive catalytic fusion reactions of
nitrogen and/or oxygen nuclei captured in air, but the range in
air of the heavy products of these reactions is $<$0.5 cm, so
that the probability of their detection after the daemon has
traversed the top polystyrene plate appears low. The same applies
to carbon and hydrogen, the components of the polystyrene, all
the more so that in their subsequent passage through the ZnS(Ag)
layer the already captured ${\rm ^{12}C}$ or ${\rm ^{1}H}$ nuclei
would not have a good chance to react outside the plate, as they
would be promoted to higher lying levels, and lost in the
preferential capture of heavier nuclei in ZnS(Ag). On passing the
7-cm gap, the daemon traverses the second polystyrene plate with
the ZnS(Ag) coating on its lower surface and enters the space
bounded by a semi-cubic sheet-metal case, which fixes the PM tube
at a distance of 22 cm from the scintillator. The presumed
phenomena occurring here are similar to those taking place after
the passage of the upper polystyrene plate, however there is more
room (and time) for their manifestation.

When moving {\it from below}, daemon enters the chamber viewed by
the second PM tube through a 0.3 mm-thick tinned iron sheet. On
exiting the Sn or Fe atom, it carries away with a high
probability its captured nucleus. On the capture, some liberated
energetic Auger electrons are able to escape from the sheet to
excite the bottom scintillator. Afterwards, in $\tau_{\rm ex}$,
the daemon releases the captured nucleus with emission of the
products of nucleon decay. After this, the daemon captures a
light nucleus in the air and emits Auger electrons, which also
excite scintillations in the bottom plate. In traversing this
plate in its motion upward, the daemon gets rid in it and in its
ZnS(Ag) layer of the light nucleus captured in the air and
recaptures a heavier ion (or nucleus) from ZnS(Ag) carrying it
into the polystyrene bulk. While being in the ZnS(Ag) $\sim$10
\mkm-size grain during $<\!10^{-9}$ s, the daemon probably has no
time to force the nucleus to emit all its radiations. So it is
doubtful that some particles shed in the polystyrene plate bulk
would be capable of reaching the ZnS(Ag) bottom layer to excite
an intensive scintillation.

When entering further the 7-cm gap between the plates or the
ZnS(Ag) upper layer, the daemon (re)captures here again such
nuclei with the emission of Auger electrons and nuclear
radiations. These latter have to be detected by the first (upper)
scintillator that triggers the oscilloscope. However keeping in
mind that some of the particles are emitted after the daemon has
penetrated well into the polystyrene bulk, one cannot be sure of
having excited a strong scintillation.

Thus, there is a variety of conceivable processes capable of
creating a scintillation at the passage of a daemon. We shall not
attempt to list them all here. It is also clear that far from all
of the weak scintillations are recorded. The strongest are
apparently the informative events initiated at a time by numerous
($\sim$10) Auger electrons and by nuclear particles. Therefore we
can lower our electronics response level and thus we have no
serious problems with numerous one-particle background event
discrimination (see Fig.1 below). In any case, however, the
response of our setup should be asymmetric with respect to the
daemon propagation direction.

\section{Some specifics of the experiment and its results}

After control tests in January-February 2000, the system was put
in round-the-clock operation in March 2000, with the total
exposure amounting to 700 h. Altogether, 
$\sim$6$\times${}${10^5}$ oscilloscope triggering events have
been recorded, only $\sim${}$10^4$ of which contained a signal on
the second trace and were entered into the computer. About 2/3
`single' triggers contain a tailing signal typical of HPS. These
signals are most likely due to radioactive background decays. The
remaining single triggers are of the NLS type. The double events
are primarily NLS signals occurring without any time delay and
coinciding in shape (delays $\le$0.2 \mks). Sometimes these
signals appear simultaneously even in all four modules. We assign
such events to cosmic rays and neglect them. Very infrequently,
once in only about twenty--thirty of all the events recorded, one
of the two signals has the HPS characteristics. Interestingly,
events with two and more significant signals in one channel are
very rare. Therefore, while basing on the above scenarios of the
events accompanying a daemon traversal of our system, one could
expect numerous signals on the same oscilloscopic trace, we began
with analyzing the sweeps containing one signal only.

The number of events with {\it shifted} signals in both traces
recorded during the month is 413. In the case of purely
non-correlated stochastic generation of signals, no statistically
significant clusters of events should appear in the time
distribution of second-trace signals.

This experiment was aimed at detecting objects moving with
velocities $\sim$35--50 ${\rm km\, s^{-1}}$.  We expected signals
with positive shifts ${\rm \Delta}t \approx 1.5-2.0$ \mks\ or
slightly more. As always, reality introduced substantial
corrections into our speculations. This relates to both the
daemon population discovered by us and the details of the events
(see Sec.2 above) accompanying daemon passage through our system.

Fig.1 shows an $N({\rm\Delta}t)$ distribution of second-trace
signals in the time ${\rm\Delta}t$ of their shift relative to the
onset of the triggering signal on the first oscilloscopic trace.
This is a fairly asymmetric and non-monotonic distribution. By
the $\chi^2$ criterion, the C.L. of this being not a
${\rm\Delta}t$-independent distribution is not less than 99.8\%.
First of all, there is no noticeable clustering of events near a
few \mks, as we expected to be based on the hypothesis of existence
of the SEECHO population. The main feature is a maximum in the
region of +30 \mks\ for an average 41.3 event/bin level. It
contains 62 events, which exceeds by a factor of 3.22 the
statistically allowable scatter $\sigma = \sqrt{41.3}\approx
6.43$. Accepting only an excess of 1.22 over 2$\sigma$, this
yields seven to eight events which can be assigned to
nuclear-active objects crossing the detector. For its area of 1
${\rm m^2}$ and an exposure time of $2.5\times 10^6$ s, this
amounts to a total flux 
$f_\oplus\approx 3\times 10^{-6}\ {\rm m^{-2}s^{-1}}$.

Our initial goal was detection of the DM objects, not their flux
measurement. We possibly are not able now to reveal all the
daemons traversing our system due to their partial poisoning with
heavy nuclei, etc. So the real flux of daemons can reach
$f_\oplus\approx 5\times 10^{-5}\ {\rm m^{-2}s^{-1}}$. An absence
of a symmetric to ${\rm\Delta}t\approx+30$ \mks\ negative feature
in $N({\rm\Delta}t)$ distribution can be attributed to
long-lasting Sn nucleus poisoning of daemons moving from below.
The various cross-checks did not reveal any noticeable systematic
errors in the operation of our simple measuring instrumentation.

\section{An attempt at interpreting the results}

\begin{figure}
\centerline{\psfig{figure=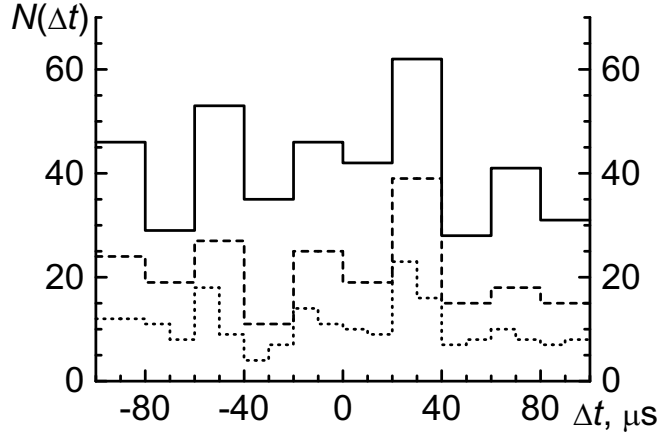,width=8.8cm}}
\caption[]{Distribution $N({\rm\Delta}t)$ of pair scintillation
events on their time shift (relative to the upper channel
events). (- - -) Similar distribution for the HPS (heavy-particle
scintillation) type events only. (....) The 10-\mks\ bin HPS
distribution.}
\label{fig1}
\end{figure}

An analysis of the $N({\rm\Delta}t)$ distribution displayed in
Fig.1 permits certain conclusions both on the nature of the agent
responsible for this distribution and on the character of its
interaction with matter. While one cannot rule out a possibility
of other interpretations, we shall try to treat the results
within the daemon hypothesis. We note immediately that when
compared with the 7 cm interplate distance, the position of the
maximum on the time axis (+30 \mks) indicates a fairly low
velocity of the maximum-forming objects. This velocity is about
2--3 ${\rm km\,s^{-1}}$ only. Taking into account the possible
slope of the trajectories could double the velocity at most.
Initially, we attempted to explain such a low velocity as due to
its characterizing the population in geocentric orbits
intercepting the Earth surface, which was captured from the one
moving in SEECHOs \cite{dr00c,dr01}.

This interpretation meets, however, with a difficulty pointed out
as far back as 1965 by Markov \cite{mar65}. The fact is that
despite their giant penetrating ability the daemons moving with a
velocity of $\sim$10 ${\rm km\, s^{-1}}$ can traverse the Earth
only $\sim${}$10^2$--$10^3$ times. Their buildup in the Earth during
4.5 Byr and interaction with the material (even if only the
proton decay with an energy release $\sim$1 BeV during 
$\tau_{\rm ex}=10^{-5}$ s is taken into account, see below)
should bring about an energy dissipation corresponding to a heat
flux of $\sim$2$\cdot 10^5$ ${\rm erg\,cm^{-2}s^{-1}}$. This
figure exceeds at least by four orders of magnitude the flux
emanating from the Earth's mantle (10 
${\rm erg\,cm^{-2}s^{-1}}$).

The assumption of the daemon velocities ranging widely in
magnitude and directions comes also in conflict with the
narrowness of the maximum ($20<{\rm\Delta}t<40$ \mks) in the
$N({\rm\Delta}t)$ distribution. In view of the fact that this
distribution was obtained from sweeps containing only one signal,
it appears that we were too optimistic by assuming that the
daemon frees itself of the captured heavy nucleus and recovers
the catalytic properties in as short a time as 
$\tau_{\rm ex}=0.1$--1 \mks, which is shorter than the time
required to cross the 7-cm gap between the plates. As already
mentioned, an analysis of the solar energetics \cite{dr00a,dr00b}
allows considerably larger values, up to
$\tau_{\rm ex}=10^{-5}$--$10^{-4}$ s. We have thus to admit that
our starting scenario of a possible sequence of events (see Sec.
2) initiated by a daemon traversal of the detector contains
excess or weakly revealing processes (capture of nuclei from the
air as a result, say, of the daemon shedding heavy nuclei during
the time it crosses the case etc.). One could go still further
and assume that the capture by the daemon of a new nucleus, which
is accompanied by ejection of a large number of Auger electrons
and nuclear particles, occurs in our small system (10--50 cm)
only when a new nucleus is recaptured during the entry into a
material with a larger atomic weight. In this case, all pieces of
the puzzle fall into place, and the sequence of the events
accompanying the daemon traversal of the system looks somewhat
differently. To begin with, on having crossed the roof (Fe, Zn)
and the floors (Mg, Al, Si, Fe, O) of our building, the daemon
reduces the mass of the captured nucleus in $10^{-5}$--$10^{-4}$
s to such an extent that, on penetrating into the top ZnS(Ag)
layer, it can already capture a S or Zn nucleus (this is possibly
the only time where we directly invoke the hypothesis of the
decay of a daemon-containing nucleon, so that for $Z_{\rm n}=26$
$\tau_{\rm ex}\approx 10$--100 \mks, a figure that still can be
reconciled with the estimates based on solar energetics
\cite{dr00a,dr00b}). The Auger electrons, nucleons, and their
clusters ejected in the process excite HPSs in the scintillator.
Not having enough time to reduce substantially the mass of the
nucleus captured here, the `poisoned' daemon enters, 7 cm
thereafter, the bottom scintillator, but it does not excite it.
The NLS excitation in the bottom scintillator is triggered by the
energetic $\sim$0.1--1 MeV long-range Auger electrons, which are
ejected at the recapture of still heavier Sn or Fe nuclei, when
the daemon reaches the lower lid of the tinned-iron case. (If it
leaves the lower part of the case through its side wall most
of the recapture Auger electrons released as the daemon traverses
the material move almost perpendicular to this wall, i.e.,
parallel to the ZnS(Ag) layer. As a result, most of the electrons
ejected from the side wall do not enter the scintillator and,
thus, will not excite a strong scintillation.) The above
reasoning suggests also that the side walls of the upper half of
the case, while "poisoning" the daemons crossing them by Sn or Fe
nuclei, leave only a solid angle of $4{\rm\pi}/6$ ster for the
daemons to pass freely into the system (through the black paper).
It thus becomes clear that the distance to be taken into account
is the separation of 29 cm between the top scintillator and the
lower case lid, and the angular spread of trajectories of the
detectable daemons is limited by a solid angle of $\sim$2 ster.
These factors are responsible for the narrowness of the maximum
at 30 \mks\ and yield 10--15 ${\rm km\,s^{-1}}$ for the velocity.
We immediately see that the latter value is in a good agreement
with the velocity of the objects falling on the Earth from
NEACHOs. It appears that they are possibly transferred here
through perturbations by the Earth (including traversal of its
material) from the population in SEECHOs, which was captured by
the Sun with the help of the Earth, and which was the initial
target of our search. The concentration of the objects found by
us in NEACHOs is determined by the balance between their capture
and subsequent ejection 1--10 Myr later, through gravitational
perturbations by the Earth, into the region of other planets'
action (as well as out of the Solar system altogether). Because
of these orbits being close to that of the Earth, the flux of the
particles of this population through the Earth should exceed the
flux of the NEACHO-population replenishing SEECHO population,
which is exactly what is observed.

An object impacting with this velocity cannot excite atoms. The
fact that oscilloscopic traces typical of data presented in Fig.1
exhibit only one pulse shifted relative to the primary one, and
that there is no pulse paired with the trigger (i.e. without a
shift in time) suggests that the radiation emitted in interaction
of the daemon with matter has a low penetrating power.
Polystyrene 4 mm thick stops it completely. If these are
electrons, their energy is $<$1 MeV.

In conclusion, consider the important information that can be
gained by using for the $N({\rm\Delta}t)$ plot only the events
containing HPSs in the first channel (Fig. 1). In accordance with
the above analysis of the sequence of the events initiated by
daemon traversal of our detector, most of these events
($\sim$80\%) have NLSs in the second channel; this is a
consequence of the recapture by daemons of nuclei in the bottom
lid accompanied by the ejection of electrons, with most of the
latter, after passing a distance of 22 cm in the air, impinge on
the bottom ZnS(Ag) screen. While this distribution consists of
212 events only, the C.L. of its being different from
$N({\rm\Delta}t)$ = const, calculated by the $\chi^2$ criterion,
becomes 99.9\%. All the 39 events responsible for the maximum
near +30 \mks\ are concentrated in this distribution. This
maximum exceeds the mean level by a factor 3.86$\sigma$. Fig.1
displays also an HPS distribution for 10-\mks\ wide bins. The
absence of features in the remaining NLS distribution provides
one more argument for the events of interest to us here not being
the result of interference or regular instrumental malfunctions.

\section{Discussion and main conclusions}

One usually searches for DM objects with $V=200$--300 
${\rm km\, s^{-1}}$, which populate the Galactic halo. We were
the first to look for a much denser near-Sun population
\cite{dr97,dr00a}. In choosing a method capable of their
detection, we counted primarily on the specific activity of the
objects of this population at the nuclear and subnuclear levels
rather than on the purely collisional interactions with particles
of the matter, as this is done, for instance, in experimental
search for much less massive (and more numerous) hypothetical
WIMPs. In these and similar experiments, the discrimination used
to reveal the expected signals is so tight \cite{ber00} as to
practically exclude the possibility of detecting fairly
infrequent but very energetic daemon signals, which become
manifest in specific conditions (for instance, when entering a
material with a high atomic number).

Disregarding some details in the possible interpretation of the
results of our experiments, which at first glance might appear
very simple, it can be maintained that we have detected at a
${\rm C.L.}\gtrsim 99.9$\% an indication of the existence of some
highly penetrative nuclear-active cosmic radiation whose objects
move with low astronomical velocities 
($V=10$--15 ${\rm km\,s^{-1}}$). They appear to have an enormous
penetrating power and are capable of passing near and through the
Earth to populate finally NEACHOs. The main indication of the
existence of this superslow radiation is the non-stochastic
component in the long-time shifts of signals of two PM tubes
viewing phosphor coatings on two parallel plates arranged one
beneath the other. This component crowds in the 20 \mks\ range.
The simplicity of our detector system, however, is only apparent,
in that in actual fact it is a result of a hard search that has
been going on for many years \cite{dr97,dr00a,dr99}. The observed
intense scintillations originate from a simultaneous ejection of
many ($\ge$10) energetic particles, and the events occurring in
the detector, as we have seen, are very complex. The order in
which they appear and the manifestation itself depend on a number
of factors which might seem to be of minor importance. Revealing
these factors and their effectively operating combinations has
become possible due to our gradually refining understanding of
the various aspects of the problem. If we had not used tinned
iron sheets for the scintillator cases, which play a role of not
passive but of the essential pieces of detector, or black paper
for their top covers, or elements of asymmetry in the
scintillator coating orientation, etc., the efficiency of daemon
detection would have dropped strongly. On the other hand, the
observed distribution with such a well pronounced, narrow shifted
peak cannot be produced, say, by neutrons thermalized somewhere
outside the detector, or by long-lived nuclear states excited,
for instance, by cosmic rays. Obviously enough, because of the
unavoidable dispersion in the neutron velocity or nuclear
deexcitation times such a distribution would be fairly difficult
to produce even deliberately. In our case, with a sweep $\pm$100
\mks\ long, these effects could at best increase slightly the
average background level.

Judging from its manifestations and properties, the discovered
radiation can be identified in a rather self-consistent manner
with daemons, i.e. hypothetical primordial Planckian objects
carrying an electric charge and moving in close to the Earth's
orbits. The primary source of this population is apparently the
DM of the Galactic disk. The intermediate stage is a SEECHO
population producing apparently a $\sim$35--50 ${\rm km\,s^{-1}}$
flux at the Earth at a level noticeably lower than 
$10^{-4}\ {\rm m^{-2}s^{-1}}$, which accounts for our not having
yet detected it. It is a search for similar objects with a local
concentration enhanced strongly by the gravitational action of
the Sun and the Earth that has initiated this experiment. Several
modified experiments are currently under way.

\section*{Acknowledgements}

The author is greatly indebted to M.V.Beloborodyy, R.O.Kurakin,
V.G.Latypov, K.A.Pelepelin for providing software and electronics
maintenance.

\end{document}